\documentclass[12pt]{article}
\begin{document}
\baselineskip=0.7cm
\renewcommand{\theequation}{\arabic{section}.\arabic{equation}}
\renewcommand{\thesection}{\arabic{section}.}
\renewcommand{\thesubsection}{\arabic{section}.\arabic{subsection}}
\makeatletter
\def\section{\@startsection{section}{1}{\z@}{-3.5ex plus -1ex minus
 -.2ex}{2.3ex plus .2ex}{\large}}
\def\subsection{\@startsection{subsection}{2}{\z@}{-3.25ex plus -1ex minus
 -.2ex}{1.5ex plus .2ex}{\normalsize\it}}
\def\appendix{
\par
\setcounter{section}{0}
\setcounter{subsection}{0}
\def\thesection{\Alph{section}}}
\newdimen\ex@
\ex@-.0925ex
\def\dddot#1{\raise\ex@\hbox{${\mathop{#1}\limits^{
             \vbox to-4.2\ex@{\kern-\tw@\ex@\hbox{\rm...}\vss}}}$}}
\makeatother
\def\thefootnote{\fnsymbol{footnote}}
\begin{flushright}
hep-th/9903025\\
UT-Komaba/99-5\\
March 1999
\end{flushright}
\vspace{1cm}
\begin{center}
\Large
Higher-derivative terms in one-loop effective action\\
for general trajectories of D-particles\\
in Matrix theory

\vspace{1cm}
\normalsize
{\sc Yuji Okawa}
\footnote{
E-mail:\ \ okawa@hep1.c.u-tokyo.ac.jp}
\\
\vspace{0.3cm}
{\it Institute of Physics\\
University of Tokyo, Komaba, Tokyo 153-8902, Japan}

\vspace{1.3cm}
Abstract\\

\end{center}
The one-loop effective action
for general trajectories of
D-particles in Matrix theory
is calculated
in the expansion with respect to
the number of derivatives up to six,
which gives the equation of motion consistently.
The result shows that
the terms with six derivatives vanish
for straight-line trajectories,
however, they do not vanish in general.
This provides a concrete example that
non-renormalization of twelve-fermion terms
does not necessarily imply
that of six-derivative terms.

\newpage
\section{Introduction}
\setcounter{equation}{0}
To construct a consistent quantum theory including gravity
is one of the most important problems in theoretical physics today.
The most natural approach,
namely the second quantization of the metric field,
has turned out to suffer from non-renormalizability.
We have to search for another way
to describe gravity which reconciles
controllable behavior in the short-distance region
with general covariance in the low-energy region.

Matrix theory \cite{BFSS},
which was proposed as a matrix model
for M theory \cite{M-theory},
provided a novel possibility
of quantum description of gravity
based on the description of Dirichlet branes \cite{D-brane}
in terms of the super Yang-Mills theory \cite{bound-states, DKPS}.
The effective action of D-particles in Matrix theory
calculated in the loop expansion is in precise agreement
with that obtained from the eleven-dimensional
supergravity in the low-energy region
up to two loops \cite{BB,BBPT,3-body,recoil}.
However, we only understand the reason
why gravity could emerge in Matrix theory
indirectly through the superstring theory
which underlies Matrix theory.
It is desirable to understand the reason
within the framework of the super Yang-Mills theory.
The supersymmetries of the model
certainly play an important role there.
However, it is still uncertain to what extent
the effective action of Matrix theory is
constrained by them.
It would be necessary
to investigate the nature of interactions
described by Matrix theory from various viewpoints
for deeper understanding.

The one-loop effective action
of Matrix theory has been discussed from many aspects.
In particular, it was shown that
it produces interactions of the linearized supergravity
between an arbitrary pair of M-theory excitations \cite{KT}
including the effects 
of arbitrary background configurations
of the fermionic field
in Matrix theory \cite{TR}.
Besides the contribution which corresponds
to the supergravity,
the one-loop effective action of Matrix theory
contains higher-derivative corrections.
This is not unexpected
since Matrix theory is a model for M theory,
not for the supergravity: It must contain
corrections to the supergravity
in the short-distance region.
The higher-derivative corrections at one loop
can be considered as
a part of such corrections.
The primary purpose of the present paper
is to investigate these corrections at one loop.

In investigating them, we want to 
extract information on
interaction Lagrangian from the scattering phase shift,
which is the quantity
we can first obtain from Matrix-theory calculations,
with the criterion that
the Lagrangian should produce the phase shift correctly.
We first point out an ambiguity in this procedure
which is related to the difference
between the time in Matrix theory and that in the Lagrangian,
and then fix it with an assumption that
the two coincide.
This assumption ensures that
the equation of motion of D-particles \cite{recoil}
is consistently derived from the Lagrangian.
The result indicates that
the interaction Lagrangian contains
terms with six derivatives at one loop.

This observation raises an interesting question
related to the constraints imposed by
the supersymmetries.
It was shown by Paban, Sethi and Stern \cite{PSS1,PSS2}
that eight- and twelve-fermion terms
can appear only at one loop and at two loops,
respectively.
Since eight- and
twelve-fermion terms belong to
the same multiplets of the supersymmetries
as four- and six-derivative terms, respectively,
their results indicate that
the latter terms 
are strongly constrained by the supersymmetries
as well.
However,
if there really exist six-derivative terms
at one loop,
the mechanism which constrains the six-derivative terms
would be more complicated than that
for the twelve-fermion terms.

These motivate us to develop
systematic methods
to investigate higher-derivative terms
in the effective action of Matrix theory.
We explicitly calculate the one-loop effective action
for general trajectories of D-particles
in the expansion with respect to the number of derivatives
up to six
hoping that
the result provides a basis for future studies
to understand the role of the supersymmetries
and the reason why gravity could emerge in Matrix theory.

The organization of the paper is as follows.
In Section 2, after pointing out the ambiguity
in determining the interaction Lagrangian from the phase shift,
we argue that there exist six-derivative terms
at one loop under the assumption
that the time in Matrix theory
coincides with that in the Lagrangian.
We then determine the complete form of
the interaction Lagrangian up to six derivatives
by calculating the effective action for
general trajectories of D-particles in Section 3.
Section 4 is devoted to the conclusions and
discussions.

\section{Extraction of interaction Lagrangian from
scattering phase shift}
\setcounter{equation}{0}

Let us reconsider the procedure to extract
information on the interaction Lagrangian from
the scattering phase shift of D-particles
in Matrix theory.
Matrix theory is defined by the action of
the super Yang-Mills theory
dimensionally reduced from 9+1 dimensions to 0+1 dimension
\cite{BFSS}
\begin{eqnarray}
S = \int_{-\infty}^\infty dt 
\left[ \frac{1}{2} {\rm tr} D_t X^n D_t X^n 
+ \frac{1}{4} g^2 {\rm tr} [X^n,X^m] [X^n,X^m] \right. 
\nonumber \\
\left. + \frac{1}{2} {\rm tr}
( i \theta^T D_t \theta
+ g \theta^T \gamma^n [X^n,\theta]) \right],
\label{Matrix_action}
\end{eqnarray}
with
\begin{eqnarray}
D_t X^n = \partial_t X^n -ig [A,X^n], \quad
D_t \theta = \partial_t \theta -ig [A,\theta],
\end{eqnarray}
where $g$ is the Yang-Mills coupling constant and
$n,m = 1,2,\cdots,9$ stand for transverse dimensions.
$X^n_{ij}(t)$, $A_{ij}(t)$ and $\theta_{ij}(t)$ are
$N \times N$ Hermitian-matrix fields. 
Eigenvalues of the bosonic field $X^n (t)$
are interpreted as
transverse coordinates of D-particles
and the fermionic field $\theta(t)$
is an $SO(9)$ Majorana spinor
which represents spins of D-particles.
We take a representation of
$SO(9)$ gamma matrices $\gamma^n$
such that $\gamma^n$
are real and symmetric satisfying
$\{ \gamma^n,\gamma^m \} = 2\delta^{nm}$.

We will perform our computations in Euclidean formulation, 
defining the Euclidean time $\tau$ and gauge field in Euclidean time
$\tilde{A}$ as
\begin{eqnarray}
\tau = it, \quad \tilde{A} = -iA. 
\end{eqnarray}
The Euclidean action is then
\begin{eqnarray}
\tilde{S} = \int_{-\infty}^\infty d \tau
\left[ \frac{1}{2} {\rm tr} D_\tau X^n D_\tau X^n 
- \frac{1}{4} g^2 {\rm tr} [X^n,X^m] [X^n,X^m] \right. 
\nonumber \\
\left. + \frac{1}{2} {\rm tr}
( \theta^T D_\tau \theta
- g \theta^T \gamma^n [X^n,\theta]) \right],
\end{eqnarray}
with
\begin{eqnarray}
D_\tau X^n = \partial_\tau X^n -ig [\tilde{A},X^n], \quad
D_\tau \theta = \partial_\tau \theta -ig [\tilde{A},\theta].
\end{eqnarray}
We consider diagonal background configurations $B^n(\tau)$
of the bosonic field $X^n(\tau)$
\begin{eqnarray}
X^n = \frac{1}{g} B^n + Y^n, \qquad
B^n_{ij} (\tau) = \delta_{ij} r_i^n(\tau),
\label{B}
\end{eqnarray}
and use the standard  background field gauge condition
\begin{eqnarray}
-\partial_\tau \tilde{A} +i[B^n,Y^n] = 0.
\end{eqnarray}

The one-loop effective action $\tilde{\Gamma}^{(1)}$
is obtained from the functional determinant 
of the quadratic part of the action
$\tilde{S}_{(2)}$ expanded around the background (\ref{B})
after the gauge fixing
as follows:
\begin{equation}
\exp [ -\tilde{\Gamma}^{(1)} ] 
= \int DY^n D \tilde{A} D \bar{c} Dc D \theta~
\exp[-\tilde{S}_{(2)}],
\end{equation}
\begin{eqnarray}
\tilde{S}_{(2)} &=&
\int_{-\infty}^\infty d \tau
\left[
\frac{1}{2} Y_{ij}^n
( -\partial_\tau^2 + r_{ij}(\tau)^2 ) Y_{ji}^n
+ \frac{1}{2} \tilde{A}_{ij}
( -\partial_\tau^2 + r_{ij}(\tau)^2 ) \tilde{A}_{ji}
\right.
\nonumber \\ && \qquad \qquad
-2i \partial_\tau r_{ij}^n (\tau) \tilde{A}_{ij} Y_{ji}^n
+ \bar{c}_{ij}
( -\partial_\tau^2 + r_{ij}(\tau)^2 ) c_{ji}
\nonumber \\ && \qquad \qquad
\left. + \frac{1}{2} 
( \theta^T_{ij} \partial_\tau \theta_{ji}
+ \theta^{T}_{ij} r\!\!\!/_{ij}(\tau) \theta_{ji} ) 
\right],
\end{eqnarray}
where we defined
\begin{equation}
r^n_{ij}(\tau) = r^n_i(\tau) -r^n_j(\tau), \quad
r_{ij}(\tau) = \sqrt{r^n_{ij}(\tau) r^n_{ij}(\tau)}, \quad
r\!\!\!/_{ij}(\tau) = \gamma^n r^n_{ij}(\tau).
\end{equation}
After the Gauss integrations, we have
\begin{eqnarray}
\tilde{\Gamma}^{(1)} &=&
\sum_{i<j} \left[
{\rm tr} \ln \left( 1 - 4 \partial_\tau r_{ij}^n (\tau)
\frac{1}{-\partial_\tau^2 + r_{ij}(\tau)^2}
\partial_\tau r_{ij}^n (\tau)
\frac{1}{-\partial_\tau^2 + r_{ij}(\tau)^2}
\right)
\right.
\nonumber \\
&& \left. \qquad
- \frac{1}{2} {\rm Tr} \ln \left(
{\bf 1} + \partial_\tau r\!\!\!/_{ij}(\tau)
\frac{1}{-\partial_\tau^2 + r_{ij}(\tau)^2}
\right)
\right],
\label{Gamma}
\end{eqnarray}
where {\rm tr} denotes the trace over the functional space
and {\rm Tr} is that over the functional space and
the spinor indices.
Note that we do not assume that
the background (\ref{B}) satisfies the equation
of motion in deriving (\ref{Gamma})
so that we can calculate the effective action
for general trajectories of D-particles
based on (\ref{Gamma})\footnote{
Similar calculations have recently been done
in the reference \cite{HM}
in the context of the generalized conformal symmetry
\cite{JY,JKY1,JKY2}.
}.
We will concentrate on a pair of D-particles
and omit the subscripts such as $ij$ in what follows.

The one-loop effective action
for the configuration which represents
the straight-line trajectories of D-particles
\begin{eqnarray}
r^n (\tau) = v^n \tau + x^n,
\label{straight}
\end{eqnarray}
is exactly evaluated \cite{DKPS}
using the proper-time integration as
\begin{eqnarray}
\tilde{\Gamma}^{(1)} =
- \int_{-\infty}^\infty d \tau
\int_0^\infty \frac{d \sigma}{\sigma}
16 \sinh^4 \frac{\sigma v}{2}
\Delta (\sigma, \tau, \tau),
\label{Gamma1}
\end{eqnarray}
where the proper-time propagator
$\Delta (\sigma, \tau_1, \tau_2)$
is defined by
\begin{equation}
\Delta (\sigma, \tau_1, \tau_2 ) \equiv
\exp [-\sigma (-\partial_{\tau_1}^2 + r(\tau_1)^2)]
\delta ( \tau_1 - \tau_2 ),
\end{equation}
and its explicit form when $\tau_1=\tau_2=\tau$ is
\begin{eqnarray}
\Delta (\sigma, \tau, \tau) =
\sqrt{\frac{v}{2 \pi \sinh (2 \sigma v)}}
\exp \left[
- v \left( \tau + \frac{x \cdot v}{v^2} \right)^2
\tanh (\sigma v)
- \sigma \left( x^2 - \frac{(x \cdot v)^2}{v^2}
\right) \right].
\end{eqnarray}
The integration over $\tau$ is easily carried out
and that over $\sigma$ can also be performed
after expanding the integrand with respect to $v$
\begin{equation}
\tilde{\Gamma}^{(1)} =
- \frac{v^3}{b^6} + 0 \, \frac{v^5}{b^{10}} 
- \frac{3}{2} \frac{v^7}{b^{14}}
+ O(v^9),
\label{phase1}
\end{equation}
where $b$ is the impact parameter
\begin{equation}
b = \sqrt{x^2 - \frac{(x \cdot v)^2}{v^2} }.
\end{equation}

This phase shift (\ref{phase1}) precisely coincides
with the one coming from the following interaction Lagrangian
in the eikonal approximation:
\begin{equation}
{\cal L} = C_4 \frac{v^4}{r^7} + C_6 \frac{v^6}{r^{11}}
+ C_8 \frac{v^8}{r^{15}} + O(v^{10}).
\label{L1}
\end{equation}
In fact, the integration over $\tau$
after substituting the straight-line trajectory
(\ref{straight}) into (\ref{L1}) gives
\begin{equation}
\int_{-\infty}^\infty d \tau {\cal L} =
\frac{16 C_4}{15} \frac{v^3}{b^6}
+ \frac{256 C_6}{315} \frac{v^5}{b^{10}} 
+ \frac{2048 C_8}{3003} \frac{v^7}{b^{14}}
+ O(v^9).
\label{phase2}
\end{equation}
The coefficients of the terms in the Lagrangian
(\ref{L1}) can be determined
by comparing two expressions
(\ref{phase1}) and (\ref{phase2}).
Thus, we obtain the Lagrangian
\begin{equation}
{\cal L} = -\frac{15}{16} \frac{v^4}{r^7} + 0 \frac{v^6}{r^{11}}
- \frac{9009}{4096} \frac{v^8}{r^{15}} + O(v^{10}),
\label{L2}
\end{equation}
which yields the phase shift (\ref{phase1}).

However, this is not the unique Lagrangian
which gives the phase shift (\ref{phase1}).
For example, there are four possible terms
which contain six derivatives
if we allow interactions of the form $v \cdot r$:
\begin{equation}
\int_{-\infty}^\infty d \tau \left[
A \frac{v^6}{r^{11}} + B \frac{v^4 ( v \cdot r )^2}{r^{13}}
+ C \frac{v^2 ( v \cdot r )^4}{r^{15}}
+ D \frac{( v \cdot r )^6}{r^{17}}
\right].
\label{L3}
\end{equation}
Since the integration over $\tau$
after substituting the straight-line trajectory
(\ref{straight}) into (\ref{L3}) gives
\begin{equation}
\frac{256}{45045} ( 143 A + 13 B + 3 C + D )
\frac{v^5}{b^{10}},
\end{equation}
the vanishing contribution proportional to $v^5/b^{10}$
in (\ref{phase1}) only requires
a relation among the coefficients
\begin{equation}
143 A + 13 B + 3 C + D = 0.
\label{relation}
\end{equation}
The term which we took in (\ref{L2})
is only a special one
which satisfies the relation (\ref{relation}).
Thus, the criterion that
the interaction Lagrangian should produce the phase shift
(\ref{phase1}) correctly alone
cannot determine the form of the Lagrangian uniquely.
We need some assumptions or further physical inputs
to determine them.

It should be noted that this ambiguity
is related to the difference between
the time $\tau$
in the Matrix-theory effective action (\ref{Gamma1})
and that in the Lagrangian such as (\ref{L2}) or (\ref{L3}).
For example, the value of the Lagrangian (\ref{L2})
for the configuration (\ref{straight})
depends on $\tau$ only through a combination
$r(\tau) = \sqrt{v^2 \tau^2 + 2 v \cdot x \tau + x^2}$.
The $\tau$-dependence of $\tilde{\Gamma}^{(1)}$ (\ref{Gamma1}) is
different from the above form
and therefore the two $\tau$'s in $\tilde{\Gamma}^{(1)}$
(\ref{Gamma1}) and (\ref{L2}) are different
when we interpret that the phase shift (\ref{phase1})
is coming from the Lagrangian (\ref{L2}).

Now, realizing the difference between the two $\tau$'s,
a natural assumption
in determining the form of the interaction Lagrangian
is that the time $\tau$ in the interaction Lagrangian should be
equal to $\tau$ appeared in $\tilde{\Gamma}^{(1)}$.
This assumption ensures that
the equation of motion of D-particles 
calculated in the one-loop approximation in \cite{recoil}
is consistently derived from the resultant Lagrangian.
In general, it is not obvious how the equation of motion
derived from another Lagrangian such as (\ref{L2})
is related to the variational equation
$\delta \tilde{\Gamma} / \delta r^n (\tau) =0$
in Matrix theory because of the difference of the time.
This assumption seems plausible naively because
the action of Matrix theory originates in
the low-energy effective action
of D-particles in the type IIA superstring theory
in the gauge that
the parametrization of the trajectories of D-particles
is equal to the time in ten-dimensional space-time.

Under the above assumption,
the form of the Lagrangian is uniquely determined.
The integrand of $\tilde{\Gamma}^{(1)}$ (\ref{Gamma1})
should be rewritten in the form
which depends on $\tau$ through the following combinations:
\begin{eqnarray}
r(\tau)^2 &=& v^2 \tau^2 + 2 v \cdot x \tau + x^2,
\nonumber \\
v \cdot r(\tau) &=& v^2 \tau + v \cdot x.
\end{eqnarray}
It follows that $x^2$ should be replaced with
$r^2 - v^2 \tau^2 -2 v \cdot x \tau$ and
the remaining $v \cdot x$ with
$v \cdot r - v^2 \tau$.
After these replacements, $\tau$ must appear
only within $r(\tau)^2$ or $v \cdot r(\tau)$
in order for the assumption to be consistent.
It is indeed the case for $\Delta (\sigma, \tau, \tau)$
\begin{equation}
\Delta (\sigma, \tau, \tau) =
\sqrt{\frac{v}{2 \pi \sinh (2 \sigma v)}}
\exp \left[
- \frac{(v \cdot r(\tau) )^2}{v^3}
( \tanh (\sigma v) - \sigma v ) - \sigma r(\tau)^2
\right],
\label{Delta}
\end{equation}
and $\tilde{\Gamma}^{(1)}$ (\ref{Gamma1}) is
rewritten as follows:
\begin{eqnarray}
\tilde{\Gamma}^{(1)} &=&
- \int_{-\infty}^\infty d \tau
\int_0^\infty \frac{d \sigma}{\sigma}
16 \sinh^4 \frac{\sigma v}{2}
\sqrt{\frac{v}{2 \pi \sinh (2 \sigma v)}}
\nonumber \\
&& \quad \times \exp \left[
- \frac{(v \cdot r(\tau) )^2}{v^3}
( \tanh (\sigma v) - \sigma v ) - \sigma r(\tau)^2
\right].
\label{Gamma3}
\end{eqnarray}
We can obtain the explicit form of the Lagrangian
in the expansion with respect to $v$, which is
\begin{equation}
\tilde{\Gamma}^{(1)} =
\int_{-\infty}^\infty d \tau
\left[
- \frac{15}{16} \frac{v^4}{r^7}
+ \frac{315}{128} \frac{v^6}{r^{11}}
- \frac{3465}{128} \frac{v^4 (v \cdot r)^2}{r^{13}}
+ O(v^8)
\right].
\label{L4}
\end{equation}
The terms with six derivatives in (\ref{L4})
satisfy the relation (\ref{relation})
so that the phase shift (\ref{phase1})
is reproduced correctly up to $v^6$ as it should be.

The first term in (\ref{L4}) proportional to
$v^4/r^7$ is the familiar one
which corresponds to the supergravity contribution
\cite{BFSS,BB,BBPT}.
The appearance of the terms which contain six derivatives
is interesting from the aspect of
the supersymmetric constraint as mentioned in Section 1.
It was shown
that twelve-fermion terms without derivatives
only appear at two-loop effective action \cite{PSS2}
and this statement is often referred as
the non-renormalization of twelve-fermion terms.
This in particular implies that
there are no twelve-fermion terms at one loop.
Although the six-derivative terms belong to
the same multiplet of the supersymmetries
as the twelve-fermion terms
because the number of derivatives
plus half of the number of fermions is the same,
the appearance of the six-derivative terms
at one loop does not contradict
the non-renormalization of
twelve-fermion terms
if the supersymmetric completion of
the six-derivative terms is achieved
without twelve-fermion terms.
This can happen and in fact
the tree-level effective action provides
a simpler example of such cases
\begin{equation}
\tilde{\Gamma}^{(0)}
= \int_{-\infty}^\infty d \tau \left[
\frac{1}{2 g^2} v^2 + \frac{1}{2 g^2} \psi \dot{\psi}
\right],
\end{equation}
where $\psi$ is the fermionic background.
There is a two-derivative term
but no four-fermion terms.
The same thing can happen
for four- or six-derivative terms
at higher loops 
and such possibility is not excluded
by the arguments put forward so far.
Therefore, the non-renormalization
of eight- or twelve-fermion terms itself
does not ensure the non-renormalization
of four- or six-derivative terms, respectively.
If there really exist the six-derivative terms
in the one-loop effective action (\ref{L4}),
this provides a concrete example.
This argument shows the importance of the question
whether there exist six-derivative terms
in the one-loop effective action.

The expression (\ref{L4}) indicates that
there are six-derivative terms,
however, it is not conclusive yet:
The terms might be arranged to a total derivative
if there are appropriate terms which contain
acceleration.
We can present evidence in favor of the fact
that there do exist
six-derivative terms at one loop from the result of
the one-point proper function calculated
in the one-loop approximation \cite{recoil}.
The one-point proper function is nothing but
the recoil acceleration $\delta \alpha^n (\tau)$
\cite{recoil}:
\begin{eqnarray}
\delta \alpha^n (\tau) &=&
\left. \frac{\delta \tilde{\Gamma}^{(1)}}
{\delta r^n(\tau)} \right|_{r^n(\tau) = v^n \tau + x^n} 
\nonumber \\
&=&
\int_0^\infty d \sigma \left[
32 r^n (\tau) \sinh^4 \frac{\sigma v}{2}
\Delta (\sigma, \tau, \tau)
\right.
\nonumber \\
&& \left. \qquad \qquad
+ 32 \frac{v^n}{v}
\cosh \frac{\sigma v}{2} \sinh^3 \frac{\sigma v}{2}
\partial_\tau \Delta (\sigma, \tau, \tau)
\right].
\label{alpha1}
\end{eqnarray}
Using the expression of $\Delta (\sigma, \tau, \tau)$
in terms of $r(\tau)^2$ and $v \cdot r(\tau)$
(\ref{Delta}),
the explicit form of $\delta \alpha^n (\tau)$
in the expansion with respect to $v$ is given by
\begin{eqnarray}
\delta \alpha^n (\tau) &=&
\frac{105}{16} \frac{v^4 r^n}{r^9}
- \frac{105}{4} \frac{v^2 (r \cdot v) v^n}{r^9}
\nonumber \\
&& -\frac{3465}{128} \frac{v^6 r^n}{r^{13}}
+ \frac{45045}{128} \frac{v^4 (r \cdot v)^2 r^n}{r^{15}}
\nonumber \\
&& +\frac{17325}{64} \frac{v^4 (r \cdot v) v^n}{r^{13}}
- \frac{45045}{32} \frac{v^2 (r \cdot v)^3 v^n}{r^{15}}
+ O(v^7).
\label{alpha2}
\end{eqnarray}
The first two terms in (\ref{alpha2}) precisely coincide
with the contributions coming from the Euler-Lagrange
equation derived from the first term in (\ref{L4}).
The fact that there are non-vanishing terms
with six derivatives in (\ref{alpha2})
shows that the six-derivative terms in (\ref{L4})
cannot be arranged to a total derivative.

We want to comment here that
the equation of motion derived from
another Lagrangian such as (\ref{L2})
does not coincide with (\ref{alpha2})
because of the difference of the time.
It may be possible to make them coincide
if we properly redefine the time in (\ref{alpha2}).
However, the time in the other Lagrangians such as (\ref{L2})
is not simply related to that in $\tilde{\Gamma}^{(1)}$
(\ref{Gamma1}) in general.
In fact, an effective way to obtain (\ref{L2})
is the following transformation of $\tau$
after exchanging the order of integrations
between $\tau$ and $\sigma$ in $\tilde{\Gamma}^{(1)}$:
\begin{equation}
\tilde{\tau} = \sqrt{\frac{\tanh (\sigma v)}{\sigma v}}
\left( \tau + \frac{x \cdot v}{v^2} \right)
- \frac{x \cdot v}{v^2}.
\label{tau-tilde}
\end{equation}
Then, $\tilde{\Gamma}^{(1)}$ is expressed
using $\tilde{\tau}$ as
\begin{eqnarray}
\tilde{\Gamma}^{(1)} &=&
- \int_0^\infty \frac{d \sigma}{\sigma}
\int_{-\infty}^\infty d \tilde{\tau}
\sqrt{\frac{\sigma v}{\tanh (\sigma v)}}
16 \sinh^4 \frac{\sigma v}{2}
\sqrt{\frac{v}{2 \pi \sinh (2 \sigma v)}}
\exp [-\sigma r(\tilde{\tau})^2]
\nonumber \\
&=& \int_{-\infty}^\infty d \tilde{\tau}
\left[
- \int_0^\infty d \sigma
\frac{16 v \sinh^4 \frac{\sigma v}{2}}
{\sinh (\sigma v)}
\frac{e^{-\sigma r(\tilde{\tau})^2}}{\sqrt{4 \pi \sigma}}
\right],
\label{Gamma2}
\end{eqnarray}
with $r(\tilde{\tau})
= \sqrt{v^2 \tilde{\tau}^2 + 2 v \cdot x \tilde{\tau} + x^2}$.
The coefficients in (\ref{L2}) is reproduced
by expanding the integrand of (\ref{Gamma2})
with respect to $v$ and performing the integration
over $\sigma$.
The peculiar relation (\ref{tau-tilde})
which depends on $v$ and $\sigma$ enforces
a significant change of the usual interpretation that
eigenvalues of $\langle X^n(\tau) \rangle$ represent
the transverse positions of D-particles at the time $\tau$
since we have to use $\tilde{\tau}$ as time
when we adopt the Lagrangian (\ref{L2}).
It is uncertain whether such change of the interpretation
makes sense,
but we believe that our assumption that
the time in the Lagrangian coincides with
that in the effective action of Matrix theory
is the natural one which automatically ensures the consistency
between the Lagrangian and the equation of motion.

The argument presented here shows that there do exist
six-derivative terms in the one-loop effective action
$\tilde{\Gamma}^{(1)}$.
However, the calculations performed so far cannot
determine the complete form of the six-derivative terms
since the effective action for the straight-line
trajectories cannot detect terms with
second or higher derivative of coordinate.
In the next section, we will compute the effective action
for general trajectories and determine them,
which is possible under our assumption
that the time in $\tilde{\Gamma}^{(1)}$ is equal to
that in the Lagrangian.

\section{One-loop effective action
for general trajectories
in derivative expansion}
\setcounter{equation}{0}

We cannot evaluate
the one-loop effective action (\ref{Gamma})
exactly for general background configurations
so that we calculate it in the expansion with respect to
the number of derivatives up to six.
Let us introduce the following abbreviations
to simplify expressions:
\begin{eqnarray}
&& \dot{r}^n \equiv \partial_\tau r^n (\tau), \quad
\ddot{r}^n \equiv \partial_\tau^2 r^n (\tau), \quad \cdots ,
\nonumber \\
&& \Delta \equiv \frac{1}{-\partial_\tau^2 + r(\tau)^2}.
\end{eqnarray}
Then, the one-loop effective action (\ref{Gamma})
is concisely expressed as
\begin{equation}
\tilde{\Gamma}^{(1)} =
{\rm tr} \ln ( 1 - 4 \dot{r}^n \Delta \dot{r}^n \Delta )
- \frac{1}{2} {\rm Tr} \ln
( {\bf 1} + \dot{r}\!\!\!/ \Delta ).
\end{equation}
To determine the form of six-derivative terms,
we expand the effective action with respect to
the number of $\Delta$'s
and evaluate the following contributions:
\begin{equation}
\tilde{\Gamma}^{(1)} =
\tilde{\Gamma}^{(1)}_2 + \tilde{\Gamma}^{(1)}_4 
+ \tilde{\Gamma}^{(1)}_6 + O(\partial_\tau^8),
\end{equation}
where
\begin{eqnarray}
\tilde{\Gamma}^{(1)}_2 &\equiv&
- 4 {\rm tr} ( \dot{r}^n \Delta \dot{r}^n \Delta )
+ \frac{1}{4} {\rm Tr}
( \dot{r}\!\!\!/ \Delta )^2,
\\
\tilde{\Gamma}^{(1)}_4 &\equiv&
- 8 {\rm tr} ( \dot{r}^n \Delta \dot{r}^n \Delta )^2
+ \frac{1}{8} {\rm Tr}
( \dot{r}\!\!\!/ \Delta )^4,
\\
\tilde{\Gamma}^{(1)}_6 &\equiv&
- \frac{64}{3} {\rm tr}
( \dot{r}^n \Delta \dot{r}^n \Delta )^3
+ \frac{1}{12} {\rm Tr}
( \dot{r}\!\!\!/ \Delta )^6.
\end{eqnarray}

In evaluating them, it is convenient
to ``normal order'' the expressions first.
By ``normal ordering'', we mean
ordering expressions which contain
functions of $\tau$, derivatives with respect to $\tau$
and $\Delta$'s to the form $f(\tau) \partial_\tau^n \Delta^m$
using the commutation relation $[ \partial_\tau, \tau] =1$.
The following formulas are useful in the normal ordering:
\begin{eqnarray}
\left[ \partial_\tau, f(\tau) \right] &=& \dot{f}(\tau),
\\
\left[ \Delta, f(\tau) \right]
&=& \Delta ( \ddot{f}(\tau) + 2 \dot{f}(\tau) \partial_\tau )
\Delta,
\\
\left[ \Delta, \partial_\tau \right]
&=& 2 \Delta ( r \cdot \dot{r} ) \Delta.
\end{eqnarray}
Note that these commutators increase the number of derivatives
by at least one.
Then, the expressions $\Delta f(\tau)$
and $\Delta \partial_\tau$ are normal ordered as follows:
\begin{eqnarray}
\Delta f(\tau) &=& f(\tau) \Delta
+ 2 \dot{f}(\tau) \partial_\tau \Delta^2
+ \ddot{f}(\tau) ( 1 + 4 \partial_\tau^2 \Delta ) \Delta^2
\nonumber \\
&& + 4 \dot{f}(\tau) ( r \cdot \dot{r} ) \Delta^3
+ O( \partial_\tau^3 ),
\label{normal1}
\\
\Delta \partial_\tau &=&
\partial_\tau \Delta
+ 2 ( r \cdot \dot{r} ) \Delta^2
+ O( \partial_\tau^2 ).
\label{normal2}
\end{eqnarray}
We should make
a technical but important remark here.
When we count the number of derivatives,
we should not count derivatives
acting directly on $\Delta$'s
since they do not necessarily generate
derivatives of coordinate in the final form
as we will see.
For example, we should count
the second term on the right-hand side of (\ref{normal1})
as $O(\partial_\tau)$, not as $O(\partial_\tau^2)$.

Let us begin with $\tilde{\Gamma}^{(1)}_2$
and $\tilde{\Gamma}^{(1)}_6$.
$\tilde{\Gamma}^{(1)}_2$ is shown to vanish
\begin{equation}
\tilde{\Gamma}^{(1)}_2 = 0,
\label{Gamma-1-2}
\end{equation}
which follows from
\begin{equation}
\gamma_{ab}^n \gamma_{ba}^m = 16 \delta^{nm}.
\end{equation}
$\tilde{\Gamma}^{(1)}_6$ is easily normal ordered
as follows:
\begin{eqnarray}
\tilde{\Gamma}^{(1)}_6 =
- \frac{64}{3} {\rm tr} [ \dot{r}^6 \Delta^6 ]
+ \frac{1}{12} {\rm Tr} [ \dot{r}\!\!\!/^6 \Delta^6 ]
+ O( \partial_\tau^7 ).
\end{eqnarray}
Then using $a\!\!\!/ a\!\!\!/ = a^2 {\bf 1}$, we have
\begin{equation}
\tilde{\Gamma}^{(1)}_6 =
-20 {\rm tr} [ \dot{r}^6 \Delta^6 ]
+ O( \partial_\tau^7 ).
\label{Gamma-1-6}
\end{equation}
We go on to the last one, $\tilde{\Gamma}^{(1)}_4$.
It is transformed using
\begin{equation}
\gamma_{ab}^k \gamma_{bc}^\ell \gamma_{cd}^m \gamma_{da}^n
= 16 ( \delta^{k \ell} \delta^{mn} 
- \delta^{km} \delta^{\ell n} + \delta^{kn} \delta^{\ell m} ),
\end{equation}
to the following form:
\begin{eqnarray}
\tilde{\Gamma}^{(1)}_4 =
-4 {\rm tr} (\dot{r}^n \Delta \dot{r}^n \Delta
\dot{r}^m \Delta \dot{r}^m \Delta )
-2 {\rm tr} (\dot{r}^n \Delta \dot{r}^m \Delta
\dot{r}^n \Delta \dot{r}^m \Delta ).
\end{eqnarray}
Let us normal order the expression
${\rm tr} (\dot{r}^k \Delta \dot{r}^\ell \Delta
\dot{r}^m \Delta \dot{r}^n \Delta )$.
There are three $\Delta$'s to be moved
to the right.
One way of doing that is to move the three
one by one in the order that
the right one, the middle one and the left one.
Then, it is normal ordered as follows:
\begin{eqnarray}
&& {\rm tr} (\dot{r}^k \Delta \dot{r}^\ell \Delta
\dot{r}^m \Delta \dot{r}^n \Delta )
\nonumber \\
&=& {\rm tr} [
\dot{r}^k \dot{r}^\ell \dot{r}^m \dot{r}^n
\Delta^4
\nonumber \\
&& \quad + 2 \{
\dot{r}^k \dot{r}^\ell \dot{r}^m [\partial_\tau, \dot{r}^n]
+ \dot{r}^k \dot{r}^\ell [\partial_\tau, \dot{r}^m \dot{r}^n]
+ \dot{r}^k [\partial_\tau, \dot{r}^\ell \dot{r}^m \dot{r}^n]
\} \partial_\tau \Delta^5 
\nonumber \\
&& \quad+ \{
\dot{r}^k \dot{r}^\ell \dot{r}^m
[\partial_\tau, [\partial_\tau, \dot{r}^n]]
+ \dot{r}^k \dot{r}^\ell [\partial_\tau,
[\partial_\tau, \dot{r}^m \dot{r}^n]]
+ \dot{r}^k [\partial_\tau,
[\partial_\tau, \dot{r}^\ell \dot{r}^m \dot{r}^n]]
\} (1 + 4 \partial_\tau^2 \Delta ) \Delta^5
\nonumber \\
&& \quad + \{
12 \dot{r}^k \dot{r}^\ell \dot{r}^m [\partial_\tau, \dot{r}^n]
+8 \dot{r}^k \dot{r}^\ell [\partial_\tau, \dot{r}^m \dot{r}^n]
+4 \dot{r}^k [\partial_\tau, \dot{r}^\ell \dot{r}^m \dot{r}^n]
\} ( r \cdot \dot{r} ) \Delta^6
\nonumber \\
&& \quad + 4 \{
\dot{r}^k \dot{r}^\ell [\partial_\tau, \dot{r}^m
[\partial_\tau, \dot{r}^n]]
+ \dot{r}^k [\partial_\tau, \dot{r}^\ell \dot{r}^m
[\partial_\tau, \dot{r}^n]]
+ \dot{r}^k [\partial_\tau, \dot{r}^\ell
[\partial_\tau, \dot{r}^m \dot{r}^n]]
\} \partial_\tau^2 \Delta^6 
\nonumber \\
&& \quad + O(\partial_\tau^7) 
].
\label{formula}
\end{eqnarray}
This is one of possible normal-ordered expressions.
If we perform the normal ordering in another way,
the result will apparently differ from this one.
However, the difference is an integral of
a total derivative and hence it does not affect
the final result.
The normal-ordered expression of 
$\tilde{\Gamma}^{(1)}_4$ is easily derived using the formula
(\ref{formula}), which is
\begin{eqnarray}
\tilde{\Gamma}^{(1)}_4 &=&
{\rm tr} [
-6 \dot{r}^4 \Delta^4
-72 \dot{r}^2 ( \dot{r} \cdot \ddot{r} ) \partial_\tau \Delta^5
-240 \dot{r}^2 ( \dot{r} \cdot \ddot{r} ) ( r \cdot \dot{r} ) \Delta^6
\nonumber \\
&& \quad + \{
-36 \dot{r}^2 ( \dot{r} \cdot \dddot{r} )
-20 \dot{r}^2 \ddot{r}^2
-28 ( \dot{r} \cdot \ddot{r} )^2
\} \Delta^5
\nonumber \\
&& \quad + \{
-240 \dot{r}^2 ( \dot{r} \cdot \dddot{r} )
-160 \dot{r}^2 \ddot{r}^2
-200 ( \dot{r} \cdot \ddot{r} )^2
\} \partial_\tau^2 \Delta^6
+ O(\partial_\tau^7)
].
\label{Gamma-1-4}
\end{eqnarray}

Next thing which we have to do is
to evaluate the quantity
$\langle \tau_1 | \Delta^n | \tau_2 \rangle$.
It is expressed in the proper-time representation as
\begin{eqnarray}
\langle \tau_1 | \Delta^n | \tau_2 \rangle
&=& \frac{1}{[-\partial_{\tau_1}^2 + r(\tau_1)^2 ]^n}
\delta( \tau_1 - \tau_2 )
\nonumber \\
&=& \frac{1}{\Gamma (n)}
\int_0^\infty d \sigma \sigma^{n-1}
e^{-\sigma [-\partial_{\tau_1}^2 + r(\tau_1)^2 ]}
\delta( \tau_1 - \tau_2 ).
\end{eqnarray}
We will arrange the operator
$\exp [-\sigma (-\partial_{\tau}^2 + r(\tau)^2 )]$
in the following form:
\begin{eqnarray}
e^{-\sigma [-\partial_{\tau}^2 + r(\tau)^2 ]}
= X e^{-\sigma r(\tau)^2} e^{\sigma \partial_{\tau}^2},
\end{eqnarray}
where an operator $X$ is defined by
\begin{eqnarray}
X &\equiv& e^{-\sigma [-\partial_{\tau}^2 + r(\tau)^2 ]}
e^{-\sigma \partial_{\tau}^2} e^{\sigma r(\tau)^2}
\nonumber \\
&=& e^{B+A} e^{-B} e^{-A},
\end{eqnarray}
with $A \equiv -\sigma r(\tau)^2$
and $B \equiv \sigma \partial_{\tau}^2$.
Using the Baker-Campbell-Hausdorff's formula twice
\begin{equation}
e^X e^Y = \exp \left\{
X + Y + \frac{1}{2} [X,Y]
+ \frac{1}{12}([X,[X,Y]]+[Y,[Y,X]]) + \cdots
\right\},
\end{equation}
$X$ can be arranged to the following form:
\begin{eqnarray}
X &=& \exp \left[
- \frac{1}{2} [A,B] - \frac{1}{3} [A,[A,B]]
- \frac{1}{6} [B,[A,B]] + O(\partial_\tau^3)
\right]
\nonumber \\
&=& \exp \left[
- \sigma^2 ( \dot{r}^2 + r \cdot \ddot{r} )
- 2 \sigma^2 ( r \cdot \dot{r} ) \partial_\tau
\right.
\nonumber \\
&& \left. \qquad - \frac{8}{3} \sigma^3 ( r \cdot \dot{r} )^2
- \frac{4}{3} \sigma^3 ( \dot{r}^2 + r \cdot \ddot{r} )
\partial_\tau^2
+ O(\partial_\tau^3)
\right].
\end{eqnarray}
The terms of the form
$[A, [A, \cdots [A,B] \cdots ]]$
could potentially contribute to $O(\partial_\tau^2)$
but they did not since
$[A, [A, [A,B]]]=0$.
Note also that
the remark on the counting of the number of derivatives
which we made below (\ref{normal2})
should be applied to the operators $\partial_\tau$
here as well because of the same reason.
The operator
$\exp [-\sigma (-\partial_{\tau}^2 + r(\tau)^2 )]$
is finally expressed as
\begin{eqnarray}
e^{-\sigma [-\partial_{\tau}^2 + r(\tau)^2 ]}
&=& \left[
1 - \sigma^2 ( \dot{r}^2 + r \cdot \ddot{r} )
- 2 \sigma^2 ( r \cdot \dot{r} ) \partial_\tau
- \frac{8}{3} \sigma^3 ( r \cdot \dot{r} )^2
\right.
\nonumber \\
&& \quad \left. 
- \frac{4}{3} \sigma^3 ( \dot{r}^2 + r \cdot \ddot{r} )
\partial_\tau^2
+ 2 \sigma^4 ( r \cdot \dot{r} )^2 \partial_\tau^2
+ O(\partial_\tau^3)
\right]
e^{-\sigma r(\tau)^2} e^{\sigma \partial_{\tau}^2}.
\end{eqnarray}
Now the quantity
$\langle \tau_1 | \Delta^n | \tau_2 \rangle$
is easily calculated based on the expression
\begin{equation}
\langle \tau_1 | e^{\sigma \partial_\tau^2} | \tau_2 \rangle
= \frac{1}{\sqrt{4 \pi \sigma}}
\exp \left[ -\frac{1}{4 \sigma} (\tau_1 - \tau_2)^2 \right].
\end{equation}
We need the explicit forms of
$\langle \tau | f(\tau) \Delta^4 | \tau \rangle$,
$\langle \tau | f(\tau) \Delta^5 | \tau \rangle$,
$\langle \tau | f(\tau) \Delta^6 | \tau \rangle$,
$\langle \tau | f(\tau) \partial_\tau \Delta^5 | \tau \rangle$
and $\langle \tau | f(\tau) \partial_\tau^2 \Delta^6
| \tau \rangle$ where $f(\tau)$ is an arbitrary function:
\begin{eqnarray}
&& \langle \tau | f(\tau) \Delta^4 | \tau \rangle
\nonumber \\
&& = \frac{1}{6} \int_0^\infty d \sigma \sigma^3
f(\tau) \left[
1 - \frac{1}{3} \sigma^2 (\dot{r}^2 + r \cdot \ddot{r} )
+ \frac{1}{3} \sigma^3 ( r \cdot \dot{r} )^2
+ O(\partial_\tau^3)
\right]
\frac{e^{-\sigma r(\tau)^2}}{\sqrt{4 \pi \sigma}},
\\
&& \langle \tau | f(\tau) \Delta^5 | \tau \rangle
= \frac{1}{24} \int_0^\infty d \sigma \sigma^4
f(\tau) \left[
1 + O(\partial_\tau)
\right]
\frac{e^{-\sigma r(\tau)^2}}{\sqrt{4 \pi \sigma}},
\\
&& \langle \tau | f(\tau) \Delta^6 | \tau \rangle
= \frac{1}{120} \int_0^\infty d \sigma \sigma^5
f(\tau) \left[
1 + O(\partial_\tau)
\right]
\frac{e^{-\sigma r(\tau)^2}}{\sqrt{4 \pi \sigma}},
\\
&& \langle \tau | f(\tau) \partial_\tau \Delta^5 | \tau \rangle
= \frac{1}{24} \int_0^\infty d \sigma \sigma^4
f(\tau) \left[
-\sigma ( r \cdot \dot{r} ) + O(\partial_\tau^2)
\right]
\frac{e^{-\sigma r(\tau)^2}}{\sqrt{4 \pi \sigma}},
\\
&& \langle \tau | f(\tau) \partial_\tau^2 \Delta^6 | \tau \rangle
= \frac{1}{120} \int_0^\infty d \sigma 
f(\tau) \left[
-\frac{1}{2} \sigma^4 + O(\partial_\tau)
\right]
\frac{e^{-\sigma r(\tau)^2}}{\sqrt{4 \pi \sigma}}.
\label{example}
\end{eqnarray}
The last of the formulas (\ref{example}) is
an example of the fact that
derivatives acting directly on $\Delta$'s
do not necessarily generate derivatives of
coordinate as we mentioned before.
The final form of the one-loop effective action
$\tilde{\Gamma}^{(1)}$ up to six derivatives is
obtained by applying these formulas
to the normal-ordered expressions
(\ref{Gamma-1-2}), (\ref{Gamma-1-6}) and (\ref{Gamma-1-4}). 
The result simplified by a partial integration is
\begin{eqnarray}
\tilde{\Gamma}^{(1)} =
\int_{-\infty}^\infty d \tau
\int_0^\infty d \sigma
\left[
- \sigma^3 \dot{r}^4
+ \frac{1}{6} \sigma^5 \dot{r}^6
+ \frac{1}{3} \sigma^5 \dot{r}^4 ( r \cdot \ddot{r} )
- \frac{1}{3} \sigma^6 \dot{r}^4 ( r \cdot \dot{r} )^2
\right.
\nonumber \\
+ \left. \frac{1}{3} \sigma^4 \dot{r}^2 \ddot{r}^2
+ \frac{2}{3} \sigma^4 ( \dot{r} \cdot \ddot{r} )^2
+ O(\partial_\tau^7)
\right]
\frac{e^{-\sigma r(\tau)^2}}{\sqrt{4 \pi \sigma}},
\label{Gamma-F1}
\end{eqnarray}
or after integrating over $\sigma$
\begin{eqnarray}
\tilde{\Gamma}^{(1)} = \int_{-\infty}^\infty d \tau
\left[
- \frac{15}{16} \frac{\dot{r}^4}{r^7}
+ \frac{315}{128} \frac{\dot{r}^6}{r^{11}}
+ \frac{315}{64} \frac{\dot{r}^4 (r \cdot \ddot{r})}{r^{11}}
- \frac{3465}{128} \frac{\dot{r}^4 (r \cdot \dot{r})^2}{r^{13}}
\right.
\nonumber \\
+ \left. \frac{35}{32} \frac{\dot{r}^2 \ddot{r}^2}{r^9}
+ \frac{35}{16} \frac{(\dot{r} \cdot \ddot{r})^2}{r^9}
+ O(\partial_\tau^7)
\right].
\label{Gamma-F2}
\end{eqnarray}
The expressions (\ref{Gamma-F1}) and (\ref{Gamma-F2})
are our final results for the interaction Lagrangian.
There are no four-derivative terms other than
the one which is already found
in the effective action for straight-line trajectories
(\ref{L4}).
Thus, this term is valid not only for straight-line
trajectories but also general ones.
This result is consistent with that in \cite{KT,TR}
where the effective action for arbitrary background
configurations (not restricted to diagonal ones) is calculated,
or that in \cite{HM}
where four-derivative terms are determined
by calculating the effective action for particular trajectories.
The complete form of
six-derivative terms is our new result.
It is not difficult to show that
this expression cannot be expressed as
a total derivative.
As a check of our calculations, we confirmed that
the Euler-Lagrange equation
derived from this Lagrangian precisely produces
the recoil acceleration (\ref{alpha2})
where the third term on the right-hand side of (\ref{Gamma-F2})
which contains acceleration does contribute to (\ref{alpha2}).

\section{Conclusions and discussions}
\setcounter{equation}{0}

We calculated the one-loop effective action
for general trajectories of
D-particles in Matrix theory
in the expansion with respect to
the number of derivatives up to six.
We determined the form of
the interaction Lagrangian
under the assumption that
the time in Matrix theory coincides
with that in the Lagrangian
which ensures that
the equation of motion is consistently
derived.
We found that there are
non-vanishing terms which contain six derivatives
although the value of them
at straight-line trajectories vanishes.
Our result provides a concrete example that
non-renormalization of twelve-fermion terms
does not necessarily imply
that of six-derivative terms.

It is possible to calculate higher-derivative
terms in the one-loop effective action
at any order following the method developed in Section 3.
Such calculations provide predictions from Matrix theory that
M theory should have these corrections
to the supergravity approximation.
Although these interactions are understood
as the remaining $\alpha'$ corrections 
in the limit discussed in \cite{DKPS,Seiberg,Sen}
in the viewpoint of
the type IIA superstring theory,
it is interesting to identify the origin
of them in M theory.

Although we did not use the supersymmetries
manifestly in the calculations,
the result must be supersymmetric
if correctly calculated.
It is desirable to perform
the supersymmetric completion of the six-derivative terms
and see if it does not require twelve-fermion terms.

Our result indicates that
the mechanism which ensures
the non-renormalization of
terms with four or six derivatives,
if it exists,
is more complicated than
that of eight- or twelve-fermion terms
which was shown in \cite{PSS1,PSS2}.
Something more will be required to prove
the non-renormalization of four- or six-derivative terms,
in particular, in multi-body systems
as discussed in \cite{Lowe},
or in \cite{DEG}
where the possibility of the existence of six-derivative terms
in four-body scattering at three loops
is argued.
However, these arguments of course do not claim
that terms with four or six derivatives
cannot be constrained only by the supersymmetries,
at least for two-body interactions.
In fact, the method taken in \cite{PSS1,PSS2}
makes use of only small part of
the potential power of the supersymmetries.
It would be necessary to exploit more powerful methods
to constrain the effective action
using the supersymmetries.

As a possible extension of our work,
it is interesting to discuss
the influence of the higher-derivative
terms in the one-loop effective action
on the generalized conformal symmetry
proposed in \cite{JY} and developed in \cite{JKY1,JKY2,HM}.
The question whether
the conformal transformation is modified
by them seems important
since the conformal transformation
is closely related to the property
of the background metric
which is not manifest in Matrix theory.

\vspace{0.4cm}
\noindent
Acknowledgements

The author would like to thank T. Yoneya
for helpful discussions and comments and also thank
M. Ikehara, Y. Kazama and W. Taylor for valuable discussions.
The author expresses his gratitude to S. Paban
for pointing out an error in (\ref{Gamma-F1})
and the corresponding one in (\ref{Gamma-F2})
in a previous version of the present paper.
This work was supported in part
by the Japan Society for the Promotion of Science
under the Predoctoral Research Program (No. 08-4158).


\vspace{0.5cm}
\small
\begin{thebibliography}{99}
\bibitem{BFSS} T. Banks, W. Fischler, S. H. Shenker and L. Susskind,
Phys. Rev. D 55 (1997) 5112.
\bibitem{M-theory} E. Witten, Nucl. Phys. B 443 (1995) 85.
\bibitem{D-brane} J. Polchinski, Phys. Rev. Lett. 75 (1995) 4724.
\bibitem{bound-states} E. Witten, Nucl. Phys. B 460 (1996) 335.
\bibitem{DKPS} M. R. Douglas, D. Kabat, P. Pouliot and S. H. Shenker,
Nucl. Phys. B 485 (1997) 85.
\bibitem{BB} K. Becker and M. Becker, Nucl. Phys. B 506 (1997) 48.
\bibitem{BBPT} K. Becker, M. Becker, J. Polchinski and A. Tseytlin,
Phys. Rev. D 56 (1997) 3174.
\bibitem{3-body} Y. Okawa and T. Yoneya, Nucl. Phys. B 538 (1999) 67.
\bibitem{recoil} Y. Okawa and T. Yoneya, Nucl. Phys. B 541 (1999) 163.
\bibitem{KT} D. Kabat and W. Taylor, Phys. Lett. B 426 (1998) 297.
\bibitem{TR} W. Taylor and M. V. Raamsdonk, hep-th/9812239.
\bibitem{PSS1} S. Paban, S. Sethi and M. Stern,
Nucl. Phys. B 534 (1998) 137.
\bibitem{PSS2} S. Paban, S. Sethi and M. Stern,
J. High Energy Phys. 06 (1998) 012.
\bibitem{HM}  H. Hata and S. Moriyama, hep-th/9901034.
\bibitem{JY} A. Jevicki and T. Yoneya, Nucl. Phys. B 535 (1998) 335.
\bibitem{JKY1} A. Jevicki, Y. Kazama and T. Yoneya,
Phys. Rev. Lett. 81 (1998) 5072.
\bibitem{JKY2} A. Jevicki, Y. Kazama and T. Yoneya,
Phys. Rev. D 59 (1999) 066001.
\bibitem{Seiberg} N. Seiberg, Phys. Rev. Lett. 79 (1997) 3577.
\bibitem{Sen} A. Sen,  Adv. Theor. Math. Phys. 2 (1998) 51;
See also a review, A. Sen, hep-th/9802051.
\bibitem{Lowe} D. A. Lowe, J. High Energy Phys. 11 (1998) 009.
\bibitem{DEG} M. Dine, R. Echols and J. P. Gray, hep-th/9810021.
\end{thebibliography}
\end{document}